\documentclass[a4paper,fleqn]{cas-dc}

\usepackage[numbers]{natbib}
\usepackage{gensymb}
\usepackage{caption}
\usepackage{enumitem}
\usepackage{units}
\def\tsc#1{\csdef{#1}{\textsc{\lowercase{#1}}\xspace}}
\tsc{WGM}
\tsc{QE}
\tsc{EP}
\tsc{PMS}
\tsc{BEC}
\tsc{DE}
\begin{document}
\let\WriteBookmarks\relax
\def\floatpagepagefraction{1}
\def\textpagefraction{.001}
\shorttitle{On the breakup frequency of bubbles and droplets in turbulence: a compilation and evaluation of experimental data}
\shortauthors{Shijie Zhong et~al.}

\captionsetup[figure]{labelfont={bf},labelformat={default},labelsep={period},name={Fig.}}

\title [mode = title]{On the breakup frequency of bubbles and droplets in turbulence: a compilation and evaluation of experimental data} 

\tnotetext[1]{Acknowledgment is made to the Donors of ...}

\author[1]{Shijie Zhong}
\author[1]{Rui Ni}
\cormark[1]
\ead{rui.ni@jhu.edu}

\address[1]{Department of Mechanical Engineering, Johns Hopkins University, Baltimore, MD-21218, USA}

\cortext[cor1]{Corresponding author}

\begin{abstract}
The dispersed phase in liquid-liquid emulsions and air-liquid mixtures can often be fragmented into smaller sizes by the surrounding turbulent carrier phase. The critical parameter that controls this process is the breakup frequency, which is defined from the breakup kernel in the population balance equation. The breakup frequency controls how long it takes for the dispersed phase reaches the terminal size distribution for given turbulence. In this article, we try to summarize the key experimental results and compile the existing datasets under a consistent framework to find out what is the characteristic timescale of the problem and how to account for the inner density and viscosity of the dispersed phase. Furthermore, by pointing out the inconsistency of existing experimental data, the key important unsolved questions and related problems on the breakup frequency of bubbles and droplets are discussed.

\end{abstract}

%\begin{graphicalabstract}
%\includegraphics{figs/grabs.pdf}
%\end{graphicalabstract}

\begin{highlights}
\item pending
\end{highlights}


\maketitle

\section{Introduction}

The turbulent breakup of bubbles and droplets gives rise to a dispersed phase characterized by a diverse spectrum of scales, which increases the interfacial area between the two phases and can find its application in various industries, such as food \citep{haakansson2019emulsion}, pharmaceuticals \citep{gupta2016nanoemulsions}, and ocean engineering \citep{verschoof2016bubble}. 
In these industries, the size distribution of the dispersed phase is crucial for the quality of the final products and overall functionality, e.g. the emulsion stability \citep{mcclements2007critical} and two-phase mass and heat transfer \citep{graham1973drop}.

The seminar works by Kolmogorov \citep{kolmogorov1949breakage} and Hinze \citep{hinze1955fundamentals}, i.e. Kolmogorov-Hinze (KH) theory, assumed that the bubble/droplet in homogeneous isotropic turbulence could only be broken by eddies with similar sizes. The competition between the eddy stress and surface tension give rise to the turbulent Weber number, $We_t$, which is defined as,
\begin{equation}
    We_t = \frac{\rho_c (\delta_D u)^2}{\sigma/D} \approx \frac{2 \rho_c (\epsilon D)^{5/3}}{\sigma}
\end{equation}
where $\rho_c$ is the density of the continuous phase, $D$ is the bubble/droplet diameter, $\delta_D u$ is the velocity scale of eddies at the scale of bubble/droplet size (for simplicity, we will use $u_e$ instead of $\delta_D u$ in the following discussions), $\epsilon$ is the turbulent dissipation rate, and $\sigma$ is the surface tension  coefficient. It was assumed that there exists a critical Weber number, above which the breakup would occur.

While the Weber number only quantifies the effect of surface tension, the damping by the inner viscosity of bubble/droplet should be measured by another dimensionless number. As suggested by Hinze \citep{hinze1955fundamentals}, the Ohnesorge number, $Oh$, can represent the ratio of viscous force to the inertial and surface tension force. $Oh$ can be expressed as,
\begin{equation}
    Oh=\frac{\mu_d}{\sqrt{\rho_d \sigma D}}
\end{equation}
where $\mu_d$ and $\rho_d$ are the dynamic viscosity and density of the dispersed phase, respectively. Although many experiments and simulations studied the dependence of the breakup frequency on $We_t$ and $Oh$, the existing data has a large variation among themselves and it remains unclear how to compile them together and provide a well-constrained breakup frequency.

Part of the problem is that it is not even clear what is the characteristic timescale to non-dimensionalize the breakup frequency. For breakup driven by turbulence, multiple timescales coexist, including the breakup time, the eddy turnover time \citep{hinze1955fundamentals}, the viscous time \citep{levich1962physicochemical,eastwood2004breakup}, as well as the large-scale shear time, which will be elaborated in this article. Although the relation between the breakup time and the eddy turnover time is reviewed and discussed in several papers \citep{qi2020towards,qi2022fragmentation,gaylo2023fundamental}, the role of large scale motion has not been paid too much attention.

This article aims at summarizing the key experimental results and compiling the data under one framework. Furthermore, by pointing out the inconsistency within existing experimental data, some important unsolved questions on the breakup frequency of bubbles are raised and some suggestions are provided for future research.

\section{Breakup mechanisms and breakup frequency}

The Boltzmann-type equation has been widely incorporated into simulation codes to accurately predict the size distribution of polydisperse particles, bubbles and droplets in turbulence \citep{marchisio2013computational,marchisio2003quadrature,shiea2020numerical}. 
This equation describes the time evolution of the number density for bubbles/droplets of a certain size $D$ at a given position $\boldsymbol{x}$ and time $t$, $n(D,\boldsymbol{x},t)$, which is first proposed by \citet{williams1985combustion}, 
\begin{equation}\label{eqn:pbe1}
    \frac{\partial n}{\partial t}+\boldsymbol{\nabla\cdot}(\boldsymbol{v}n)=-\frac{\partial}{\partial D}(Rn)+\dot{Q_b}+\dot{Q_c}
\end{equation}
where $\boldsymbol{v}(D,\boldsymbol{x},t)$ is the advection velocity of bubbles, $R=dD/dt$ is the rate of change of bubble size due to mass dissolution, and $\dot{Q_b}$ and $\dot{Q_c}$ are the rate of change of bubble number density $n(D,\boldsymbol{x},t)$ due to breakup and coalescence, respectively. For a system that has a very low bubble concentration and negligible dissolution, $\dot{Q_c}$ and the dissolution term $\partial(Rn)/\partial D$ can be neglected. By only considering breakup \citep{martinez2010considerations}, Eq. \ref{eqn:pbe1} can be simplified as:
\begin{equation}\label{eqn:pbe}
\begin{split}
    \frac{Dn(D)}{Dt}\equiv&\frac{\partial n(D)}{\partial t}+\boldsymbol{\nabla\cdot}\left[\boldsymbol{v}n(D)\right]\\
    =&\int_D^\infty m(D_0)f(D;D_0)g(D_0)n(D_0)dD_0\\
    &-g(D)n(D)
\end{split}
\end{equation}
where the source term for the bubbles/droplets of size $D$ due to breakup of larger bubbles/droplets is denoted by the first term on the right side. 
The number of daughter bubbles/droplets generated from a mother bubble/droplet of size $D_0$ is written as $m(D_0)$. $f(D;D_0)$ is the daughter bubble/droplet size distribution given the mother bubble/droplet size as $D_0$, and $g(D)$ is the breakup frequency. 
By knowing the formulas of $f(D;D_0)$ and $g(D)$, $n(D)$ can be predicted numerically from an initial value.

While many models on the breakup frequency, $g(D_0)$, have been reviewed in the literature \citep{lasheras2002review,liao2009literature}, there is still a lack of a comprehensive compilation of experimental results. As highlighted in \citep{haakansson2020experimental}, the compilation of breakup frequency data presents significant challenges due to different experimental conditions, varying experimental methodologies, and various definitions used for quantifying breakup frequency. This complexity makes it exceedingly difficult to gather and compare datasets from different papers. To the best knowledge of the authors, the majority of the existing papers focused on fitting experimental data to specific breakup frequency models without adequately addressing the normalization process or providing thorough comparisons with other datasets \cite{martinez1999breakup,wang2003novel,vankova2007emulsification,ravichandar2023turbulent}.

\section{Measurement challenges and uncertainties}

\citet{qi2020towards} plotted several models within a similar range of the energy dissipation rates and bubbles/droplets sizes together and showed clearly that these models do not agree with one another with differences over several orders of magnitude. Since there is no first-principle equation that can be solved to calculate the breakup frequency, many phenomenological models based on the collision of bubbles/droplets with eddies that contain strong enough inertial force \citep{lehr2002}, velocity fluctuation \citep{alopaeus2002simulation} or turbulent kinetic energy \citep{coulaloglou1977description,  martinez1999breakup}, have been proposed. 
% A large number of these possible mechanisms, together with their adjustable parameters or limits of integrals, leads to results with drastically different, and even contradictory, predictions. 
Such large number of possible mechanisms and different fitting parameters in the existing models leads to inconsistent predictions.

Constraining the model parameters by experimental data is challenging because (a) for most experiments conducted in emulsion, quantities are often averaged in non-homogeneous anisotropic turbulence, such as pipe flows, stirred tanks, or homogenizers; (b) the definition of breakup frequency is not consistent across different works \citep{haakansson2020validity}: several different experimental methods have been employed to estimate the breakup frequency, and each has its own uncertainties; (c) as $We_t$ becomes small, the breakup frequency is so low that measuring this quantity relies on exposing the two-phase flows to a rather steady turbulent environment for an extended period of time; the requirement on long enough residence time with continuous monitoring of the size of the dispersed phase has not yet met in most experiments \citep{lalanne2019model}; (d) the sensitivity of the experiments to the dimensionless numbers is still unclear; in addition to $We_t$, the dimensionless numbers also include, $Oh$, for oil-water emulsion and the Reynolds number, $Re$, of the background turbulence. 
The effect of the latter one is particularly unknown; only recently, \citet{qi2022fragmentation} showed the importance of the energetic and intermittent small eddies in accelerating breakup, which may imply that turbulence with very low Reynolds numbers may not have sufficient range of scales interacting with bubbles/drops. 

The breakup frequency can be estimated either (i) indirectly from the measurement of drop size distribution or (ii) directly by counting breakup events in high-speed videos. The most precise way to use the first method is to infer the breakup frequency of the bubble/droplet with the largest size \citep{haakansson2020validity,martinez1999breakup,eastwood2004breakup,vankova2007emulsification}. As for the number density of bubble/droplet with the largest size, the integration term in Eq.\ref{eqn:pbe} can be dropped, then we can get
\begin{equation}
    n(D_{max},t) = n(D_{max},t=0) \cdot \exp{\left[-g(D_{max})t\right]}
    \label{eq:gFit}
\end{equation}
Therefore, the breakup frequency can be calculated by fitting the time evolution of $n(D_{max},t)$ with an exponential function. For other sizes, however, it is much more challenging. Since it often includes inverse processes, and the problem is ill-posed and regularization or strong assumptions have to be made to solve the problem. The most popular method is to select two models, one for breakup frequency and one for daughter drop size distribution, and then perform forward calculation of the population balance equation. The difference between the calculated results and measurements can be minimized by tuning the adjustable parameters in the models. As long as a good fit can be obtained, this procedure seems to be justified.  However, \citet{qi2020towards} showed that similar drop size distribution can be reached even with completely different models by adjusting parameters. Alternatively, back calculation methods have also been developed to estimate breakup frequency without knowing which model to use {\it{a priori}} but with additional constraints, such as the assumption of self-similarity for the size distribution over time \citep{o2010study} or a power law dependence of breakup frequency on drop volume \citep{hounslow2004population}.

The latter becomes more popular because of its directness and access to high-speed cameras with better frame rate and resolution. The breakup frequency can be estimated by using the four relationships: 
\begin{subequations}
\begin{align}
    g(D_0) &= \frac{1}{t_b(D_0)} \text{, from \citep{hanvcil1988break}} \\
    g(D_0) &= \frac{P}{\tau} \text{, from \citep{gourdon1991influence}}    \\
    g(D_0) &= \frac{P}{t_b(D_0)} \text{, from \citep{coulaloglou1977description}} \\
    g(D_0) &= \frac{n_0-n_1}{\tau n_1+\sum_i t_{b,i}}\text{, from \citep{vejrazka2018}}
    \label{eq:gDef}
\end{align}
\end{subequations}
where $t_b$ is the average breakup time; $P$ is the breakup probability, $\tau$ is the mean residence time of drops; $n_0$ is the total number of drops, $n_1$ is the number for drops that did not break, and $t_{b,i}$ denotes the breakup times of each drop. \citet{haakansson2020validity} systematically evaluated these definitions and concluded that only method (d) will lead to a correct estimation of the drop size distribution. The rest three methods only approach the true values if the breakup frequency is very high (for a and c) or very low (for b).

\section{Characteristic timescales}

The first problem that has to be addressed before compiling experimental results is to select the appropriate dimensionless variables. For breakup frequency, it is important to know the characteristic timescale of the problem. 
The eddy turnover time has also been proposed as the possible characteristic timescale 
\begin{equation}
    t_e=\epsilon^{-1/3} D^{2/3}
    \label{equ:te}
\end{equation}
This argument was employed in the bubble size spectrum produced during air entrainment and fragmentation in breaking waves \citep{garrett2000connection,deane2002scale}. In the breaking wave case, the terminal bubble size distribution measured scales as $D^{-10/3}$, which can be derived by assuming $g(D)\sim 1/t_e \sim D^{-2/3}$ \citep{deane2002scale,chan2021turbulent}. 
Besides the eddy turnover time, complex large scale motion exists in most of the practical scenarios. Then the large scale mean shear can provide another timescale as 
\begin{equation}
    t_s=1/S
    \label{equ:ts}
\end{equation}
where $S$ is the shear rate. If the inner viscosity of the disperse phase is dominant, the capillary time denoted as
\begin{equation}
    t_c=\mu_d D/\sigma
    \label{equ:tc}
\end{equation}
should also be taken into account \citep{eastwood2004breakup}. 

\subsection{Timescales in jet experiments}

\begin{figure}[pos=h]
    \centering
    \includegraphics[width=\linewidth]{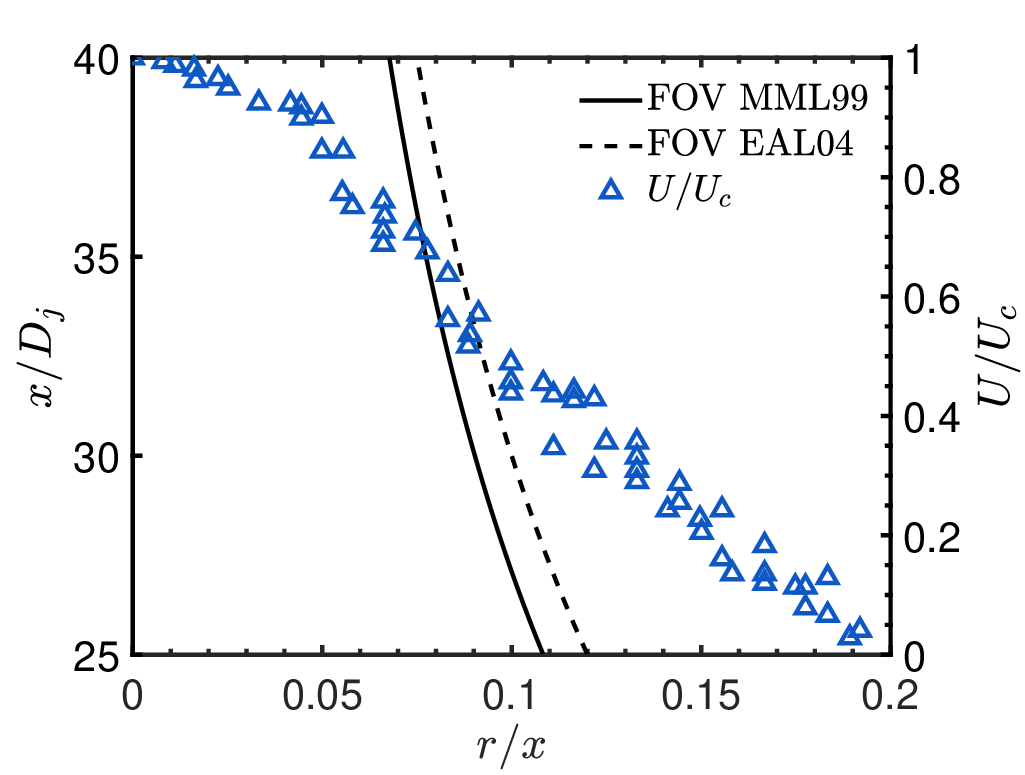}
    % \caption{Typical field of view (FOV) in turbulent jet experiments. Two datasets are compiled in this figure: MML99 \citep{martinez1999breakup}, and EAL04 \citep{eastwood2004breakup}. The right axis is the normalized mean streamwise velocity, which helps to show when high shear region goes into FOV.}
    \caption{The radial profile of the axial velocity ($U$) normalized by the centerline velocity ($U_c$) of the turbulent jet in two experiments, including MML99 \citep{martinez1999breakup} and EAL04 \citep{eastwood2004breakup}. The solid line and dashed line indicate the boundaries of the field of view (FOV) employed in these experiments. Bubbles and droplets were subjected to shear to the left of these lines.}
    \label{fig:fov}
\end{figure}

\begin{figure}
    \centering
    \includegraphics[width=\linewidth]{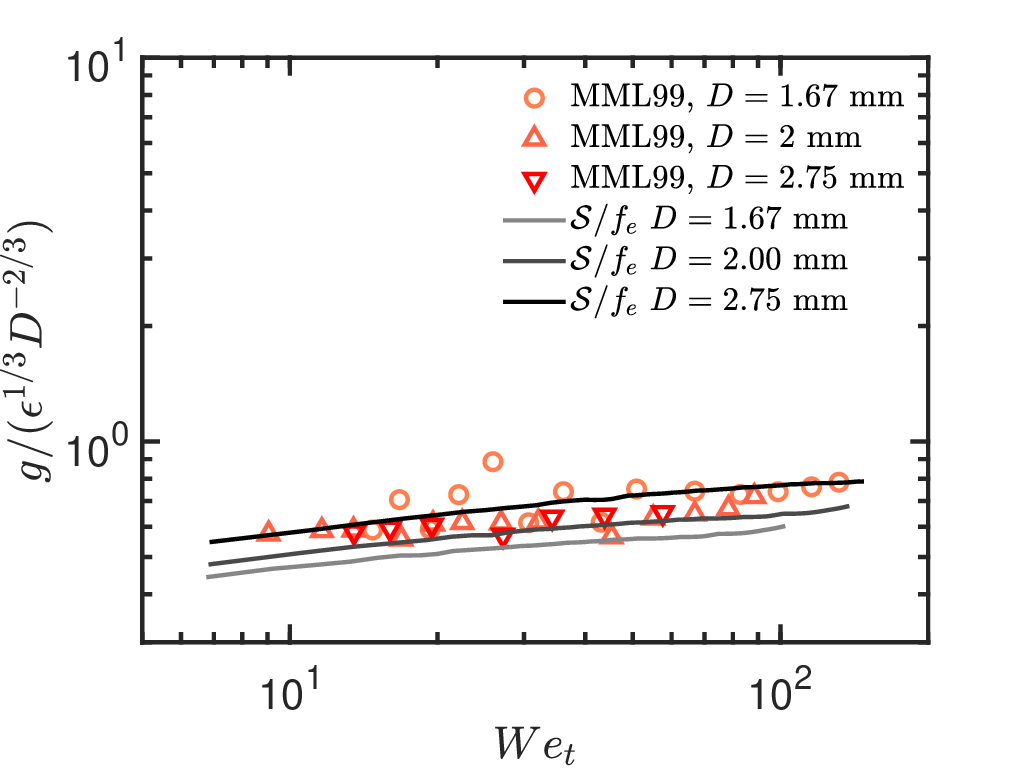}
    \caption{Normalized bubble breakup frequency as a function of $We_t$. Symbols are from the experimental data \citep{martinez1999breakup}. The solid lines represent shear rate normalized by the frequency of the bubble-sized eddies.}
    \label{fig:martinez}
\end{figure}

After summarzing the possible characteristic timescales, in this section, we discuss some experimental results that will be compiled in this paper. The first case is the turbulent jet, which is another canonical turbulent system that has well-defined characteristics.
In the studies of bubble and droplet breakup, two experiments were conducted by \citet{martinez1999breakup} and \citet{eastwood2004breakup}, respectively. A noteworthy finding in Martinez-Bazan's work was the establishment of a relationship between the breakup timescale and the eddy turnover time of the bubble size, thus corroborating the KH theory. However, the results from Eastwood's study seemed to indicate a contrasting conclusion, suggesting that the breakup timescale should be scaled with either the integral timescale or the capillary time (Eq.\ref{equ:tc}) \citep{eastwood2004breakup}. 

Despite the same turbulent jet employed, the difference in timescales reported suggests the possibility of other characteristic timescales in the system, e.g. large scale mean shear. In Fig.\ref{fig:fov}, the dashed and solid line mark the field of view at different $x$ for MML99 and EAL04, respectively. The relative width of the field of view (FOV) decreases as the streamwise distance increases. While turbulence was assumed to be homogeneous and isotropic (HIT), the presence of the mean shear within the field of view can be substantial. When $x/D_j<40$, it is shown that a non-negligible mean shear will be imposed to the bubbles/droplets when they are close to the boundary of FOV. The mean shear introduces a new timescale but such a timescale has not been considered in the literature \citep{martinez1999breakup}. The mean shear rate, i.e. $S$, is estimated as the velocity difference between the center line and the edge divided by the jet width. As shown in Fig.\ref{fig:martinez}, without any fitting parameters, the mean shear rate, calculated by averaging the shear rate over the entire jet profile, normalized by the eddy turnover frequency, $f_e = 1/t_e= \epsilon^{1/3}D^{-2/3}$, collapses well with the experimental result of the breakup frequency. Since both the magnitude and the dependence on $We_t$ of the mean shear rate are close to those of the breakup frequency, the importance of the mean shear rate cannot be disregarded. While the experimental data on the normalized break frequency decreases with size, the normalized mean shear rate increases with size. 
Nevertheless, this inconsistency is not sufficient to rule out the mean shear effect because of the limited size range adopted in experiments. 
As a result, the mean shear time could be another possible timescale in addition to bubble size eddy turnover time.

For bubbles, the inner viscosity plays a negligible role in modulating breakup dynamics. To address the viscosity effect, in the same turbulent jet setup, \citet{eastwood2004breakup} studied the breakup of droplets. In their study, four different kinds of droplets were studied, including heptane, two different types of silicone oils and olive oil. The dynamic and kinematic viscosities, as shown in Table~\ref{tab:eastwood}, cover a range over several orders of magnitude.

\begin{figure*}
    \centering
    \includegraphics[width=\linewidth]{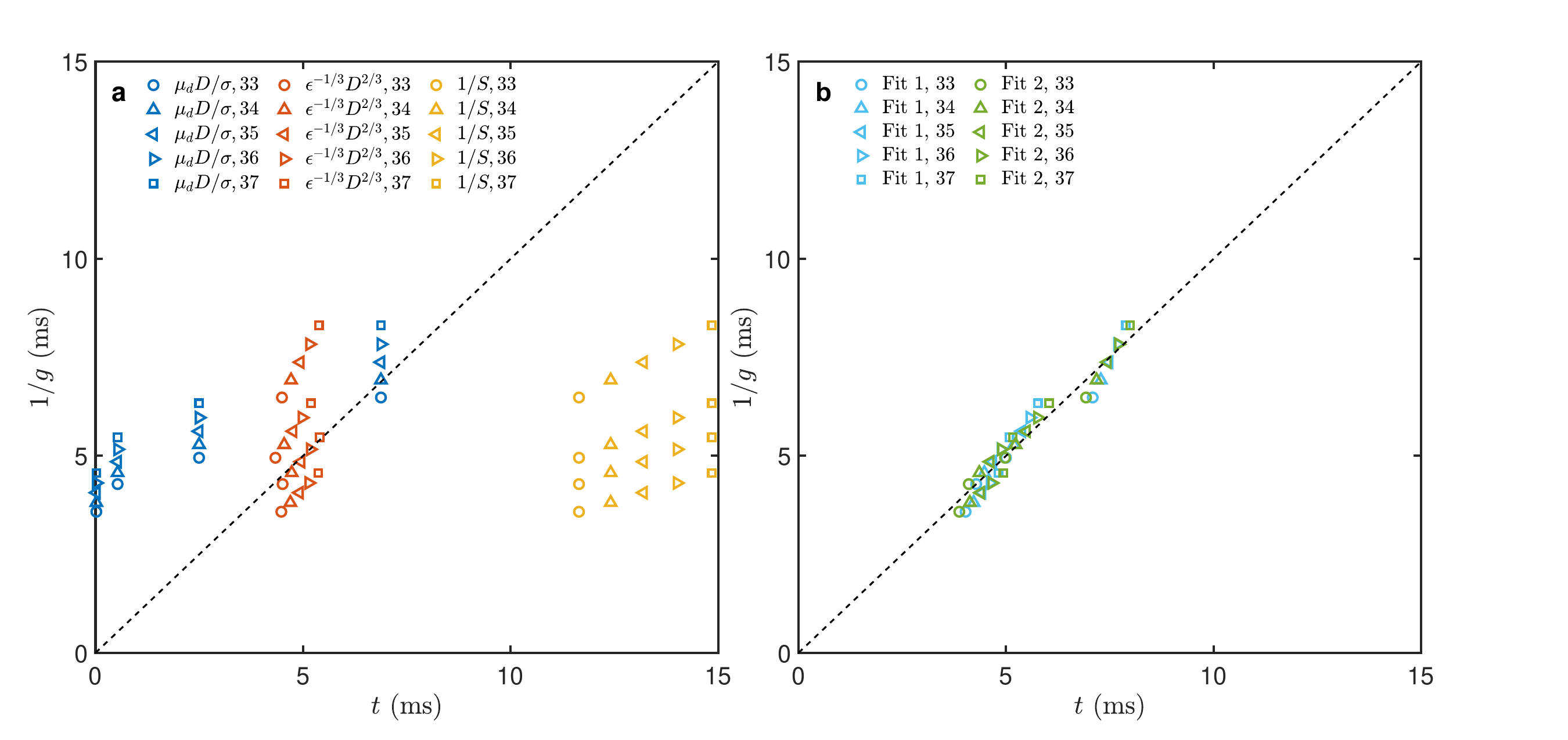}
    \caption{Breakup time, $1/g$, versus different timescales (data is retrieved from \citep{eastwood2004breakup}). The breakup time data is collected at location, $x/D_j=33,34,35,36,37$. At the same location, the values of $1/g$ from small to large represents the breakup time of heptane, 10 cSt silicone oil, 50 cSt silicone oil, and olive oil, respectively. (a) Markers in blue represent the capillary timescale; markers in orange represent bubble size eddy turnover timescale; markers in yellow represent shear timescale. (b) Fit 1 represents: $t = c_1 \cdot \mu_d D/\sigma + c_2\cdot \epsilon^{-1/3}D^{2/3}$ ($c_1=0.44,c_2=0.90$); fit 2 represents: $t = c_3\cdot \mu_d D/\sigma + c_4\cdot 1/S$ ($c_3=0.45,c_4=0.33$).}
    \label{fig:eastwood_g}
\end{figure*}

When taking the droplet viscosity into consideration, it seems that no single timescale discussed previously could explain the change of breakup frequency. As shown in panel a of Fig.\ref{fig:eastwood_g}, the breakup time, i.e. $1/g$, is plotted versus different timescales: capillary timescale, turnover time of droplet-sized eddies, and the inverse of the shear rate. 
The horizontal axis represents different timescales and vertical axis is the breakup time. Different colors represent different timescales, and different symbols represent different distances to the nozzle exit. 
If $1/g$ follows one of the timescales, the symbols with the same color would fall onto the dashed diagonal line.
% At the same location, $x/D_j=const$, the breakup times of heptane, 10 cSt silicone oil, 50 cSt silicone oil, and olive oil, are increasing, since both dynamic and kinematic viscosities are increasing. Among these three timescales, only capillary time involves droplet viscosity, so only this timescale could vary with the change of viscosity. Nevertheless, the absence of flow characteristics in the capillary timescale poses a challenge in explaining the observed changes of breakup time with location (different location has different dissipation rate). Specifically, the inability to achieve data collapse across different locations indicates the complexity of the phenomenon. 
However, it is evident that no timescale can normalize different data so they cannot collapse onto the diagonal line. It is therefore possible that the breakup process may involve multiple mechanisms, and no single timescale alone can describe the full process.

\begin{table}
\centering
\caption{Properties of droplet used in experiments of \citep{eastwood2004breakup}}
\setlength{\tabcolsep}{1mm}{
\begin{tabular}{l c c c c}\toprule
& $\rho_d$ $(kg/m^3)$ & $\mu_d$ $(Pa\cdot s)$ & $\nu_d$ $(m^2/s)$ & $\sigma$ $(N/m)$ \\ \midrule
Heptane & 684 & $5.00\times 10^{-4}$ & $7.31\times 10^{-7}$ & $4.8\times 10^{-2}$ \\ 
10 cSt Si & 936 & $9.70\times 10^{-3}$ & $1.04\times 10^{-5}$ & $3.5\times 10^{-2}$ \\
50 cSt Si & 970 & $5.09\times 10^{-2}$ & $5.25\times 10^{-5}$ & $3.7\times 10^{-2}$ \\
Olive oil & 881 & $7.19\times 10^{-2}$ & $8.16\times 10^{-5}$ & $2.0\times 10^{-2}$ \\
\bottomrule
\end{tabular}}
\label{tab:eastwood}
\end{table}

As a result, we propose that the lifetime of the bubble consists of at least two stages. In the first stage, the droplet undergoes deformation or advection by the continuous phase until it encounters a sufficiently strong eddy leading to large enough deformation. Subsequently, in the second stage, the droplet is broken due to capillary effects, which can be modulated by the inner viscosity. It is certainly plausible that the actual breakup process is far more intricate than the simplified two-stage model, particularly when multiple timescales are of similar magnitudes. In such cases, different effects may occur simultaneously, leading to a complex breakup mechanism.

When applying the two-stage model, it is crucial to consider the influences of the inner droplet viscosity and outer flow conditions, as discussed earlier. As a result, 
% the viscosity time (Eq. \ref{eqn:timeB}) is selected to account for the effects of viscosity, while either the droplet-sized eddy turnover time or the mean shear time is chosen to represent the impacts of the outer flow conditions. 
the timescale of the first stage is accounted by either the turnover time of droplet-sized eddies (Eq.\ref{equ:te}) or the inverse of mean shear rate (Eq.\ref{equ:ts}), while the capillary time (Eq.\ref{equ:tc}), is selected to account for the breakup of the neck in the second stage. 
The breakup time can therefore be expressed as the sum of two timescales, each with its corresponding weight, 
\begin{align}
    &\text{Fit 1: } t = c_1 \cdot \mu_d D/\sigma + c_2\cdot \epsilon^{-1/3}D^{2/3} \\ 
    &\text{Fit 2: } t = c_3\cdot \mu_d D/\sigma + c_4\cdot 1/S
\end{align}
By employing the linear regression, it is possible to calculate coefficients for different combinations of timescales. In panel b of Figure \ref{fig:eastwood_g}, it can be observed that both fits yield satisfactory collapse for the breakup time. Additionally, it is noteworthy that the coefficients associated with the capillary timescale ($c_1$ and $c_3$) are comparable, reinforcing the idea of the combination of two timescales.

% \begin{figure}
%     \centering
%     \includegraphics[width=\linewidth]{Fig/eastwood_tall.eps}
%     \caption{Breakup time, $1/g$, versus different timescales (data is retrieved from \citep{eastwood2004breakup}). The breakup times are collected at location, $x/D_j=33,34,35,36,37$. At the same location, the value of $1/g$ from small to large represents breakup time of heptane, 10 cSt silicone oil, 50 cSt silicone oil, and olive oil, respectively. Markers in blue represent the capillary timescale; markers in orange represent bubble size eddy turnover timescale; markers in yellow represent shear timescale.}
%     \label{fig:eastwood_tall}
% \end{figure}

% \begin{figure}
%     \centering
%     \includegraphics[width=\linewidth]{Fig/eastwood_tfit.eps}
%     \caption{Two stage model. Fit 1 represents: $t = c_1 \cdot \mu_d D/\sigma + c_2\cdot \epsilon^{-1/3}D^{2/3}$ ($c_1=0.44,c_2=0.90$); fit 2 represents: $t = c_3\cdot \mu_d D/\sigma + c_4\cdot 1/S$ ($c_3=0.45,c_4=0.33$).}
%     \label{fig:eastwood_tfit}
% \end{figure}

\subsection{Breakup in homoginizer}

In addition to larger droplets, studies on smaller droplets, e.g. nano or microemulsion, have also been conducted before. In the food industry, a device known as a homogenizer is commonly employed to break oil droplets into smaller sizes ($\sim$ 50 $\rm \mu m$) in order to achieve better emulsion. The homogenizer consists of a small channel connected to a large chamber, generating a high-speed turbulent jet that drives droplet breakup \citep{haakansson2019emulsion}. A series of comprehensive studies have been conducted on the breakup of droplets with different properties in a homogenizer \citep{vankova2007emulsificationSize,vankova2007emulsification,tcholakova2007emulsificationDaughter}. These experiments varied many parameters such as surface tension, viscosity, turbulent dissipation rate, and droplet size by orders of magnitude, resulting in a rich dataset that offers valuable insights for further analysis.

\begin{figure*}
    \centering
    \includegraphics[width=\linewidth]{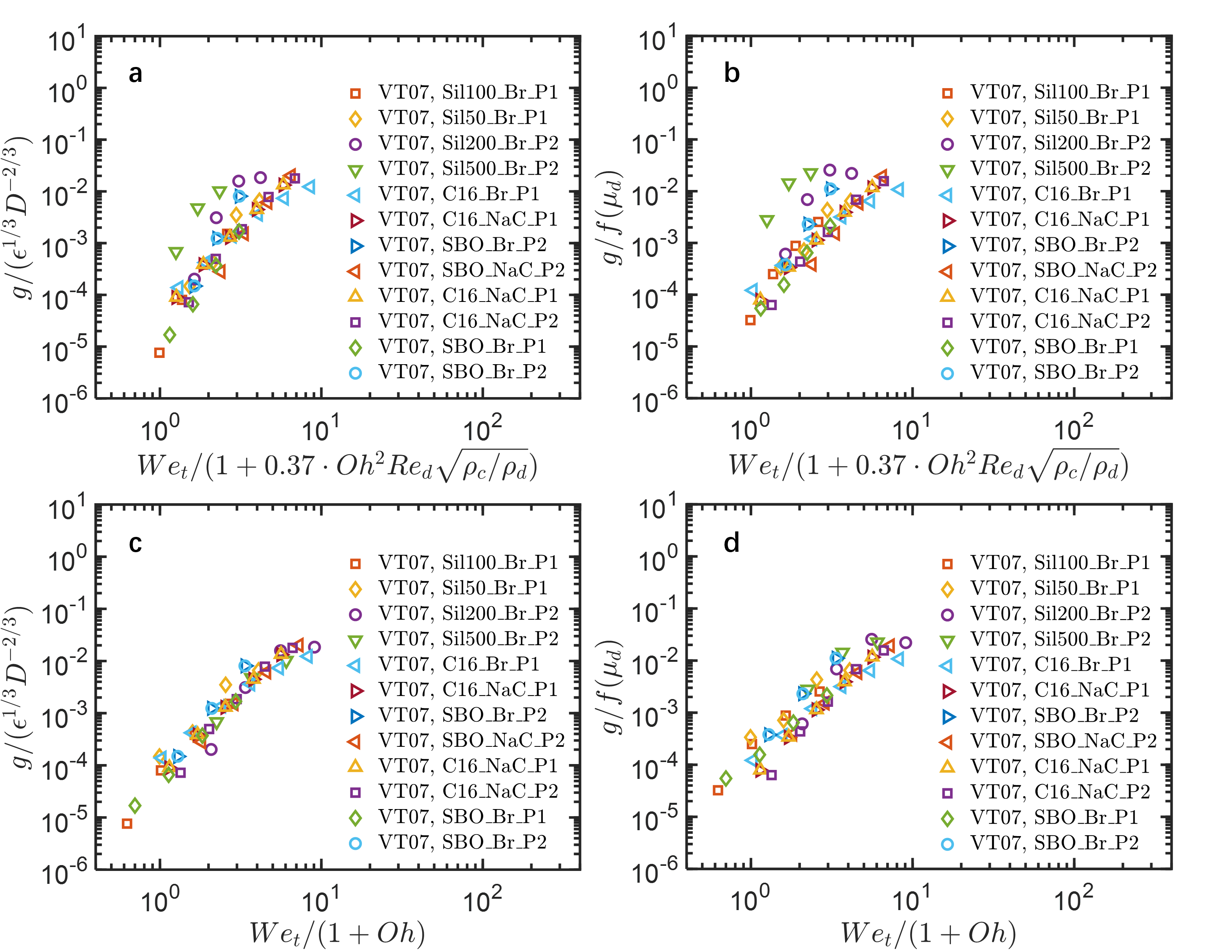}
    \caption{Normalized breakup frequency versus the effective Weber number. (a,c) the breakup frequency is normalized by the droplet-sized eddy turnover time. The horizontal axis in panel a is $We_t/(1+ c \cdot Oh^2 Re_d \sqrt{\rho_c/\rho_d} )$ with coefficient $c=0.37$ \citep{vankova2007emulsification}; while the horizontal axis in panel c is $We_t/(1+Oh)$. (b,d), the breakup frequency is normalized by Eq.\ref{eqn:fRe}.}
    \label{fig:vankova_all}
\end{figure*}

Fig.\ref{fig:vankova_all} shows the experimental results of the normalized breakup frequency as a function of different dimensionless groups by \citet{vankova2007emulsification}. To collapse all the data, they built upon the works of \citet{davies1985drop} and \citet{calabrese1986drop}. Vankova et al. assumed that droplets would not undergo breakup until the turbulent stress exceeded the combined effects of viscous and surface tension stresses,
\begin{equation}
    \rho_c\left(\varepsilon D\right)^{2 / 3} \sim
    \frac{\sigma}{D} + \frac{\mu_{d}\left(\varepsilon D\right)^{1 / 3} \sqrt{\rho_c/\rho_d}}{D}
    \label{eq:stressbalance}
\end{equation}
which can be expressed in the dimensionless form
\begin{equation}
    We_t \sim 1 + c \cdot Oh^2Re_d\sqrt{\rho_c/\rho_d}
    \label{eq:effectiveWe}
\end{equation}
where $c$ is a constant. The Reynolds number inside droplet, $Re_d$, is defined as
\begin{equation}
    Re_d = \frac{\epsilon^{1/3}D^{4/3}}{\nu_d}
\end{equation}
It was assumed that the velocity scale inside droplet is similar to the eddy velocity scale at the droplet size. 
To draw analogy to the compilation of bubble breakup frequency, the effective Weber number is defined as $We_t / (1 + c \cdot Oh^2Re_d\sqrt{\rho_c/\rho_d})$.
It basically means: when $Oh\rightarrow 0$, where viscosity effects could be ignored, the effective Weber number recovers the original Weber number; when $Oh\rightarrow \infty$, where viscosity effects dominate, the effective number asymptotically behaves like the ratio between Weber number and Ohnesorge number $We_t/Oh^2$. If the stress balance model, represented by Eq.\ref{eq:effectiveWe}, is indeed accurate, it suggests that the breakup frequency should be unaffected by surface tension when viscosity effects are dominant ($Oh \rightarrow \infty$) \citep{masbernat2022prediction}.

In order to determine an appropriate timescale for breakup frequency, \citet{vankova2007emulsification} considered two types of models based on either the droplet deformation time \citep{coulaloglou1977description} or the droplet-eddy collision frequency \citep{prince1990bubble,tsouris1994breakage,qi2023timescale}.
For the deformation time, when $Re_d<1$, the deformation is driven by the turbulent dynamic pressure and resisted by the inner viscosity. According to the derivation in \citep{levich1962physicochemical}, the corresponding timescale, i.e. $\tau$, in this case can be derived as
\begin{equation}
    \rho_c u_e^2 \sim \mu_d \frac{u_d}{D} \sim \frac{\mu_d}{\tau}  \Rightarrow \tau = \frac{\mu_d}{\rho_c u_e^2}
    \label{equ:tv_levich}
\end{equation}
where $u_d$ is the velocity scale inside the droplet. When $Re_d \geq 1$, the deformation of droplet is driven by the turbulent dynamic pressure and the inner viscosity can be ignored, so its breakup time should scale with the droplet-sized eddy turnover time. Vankova et al. added a density ratio coefficient to the eddy turnover time by following the derivation in \citep{levich1962physicochemical} as
\begin{equation}
    \frac{\rho_c u_e^2}{D} \sim \rho_d \frac{u_d}{\tau} \sim \rho_d \frac{D}{\tau^2} \Rightarrow \tau = \frac{D}{u_e}\sqrt{\frac{\rho_d}{\rho_c}}
    \label{equ:te_levich}
\end{equation}
Since $u_c\sim(\epsilon D)^{1/3}$ in turbulence, the frequency that is used to normalize the breakup frequency can be expressed as,
\begin{equation}
    f(\mu_d) = \left\{
        \begin{aligned}
            &\rho_c (\epsilon D)^{2/3}/\mu_d, \hspace{2em} Re_d < 1 \\
            &\epsilon^{1/3}D^{-2/3}\sqrt{\rho_c/\rho_d}, \ Re_d \geq 1
        \end{aligned}
    \right.
    \label{eqn:fRe}
\end{equation}

\begin{figure*}
    \centering
    \includegraphics[width=\linewidth]{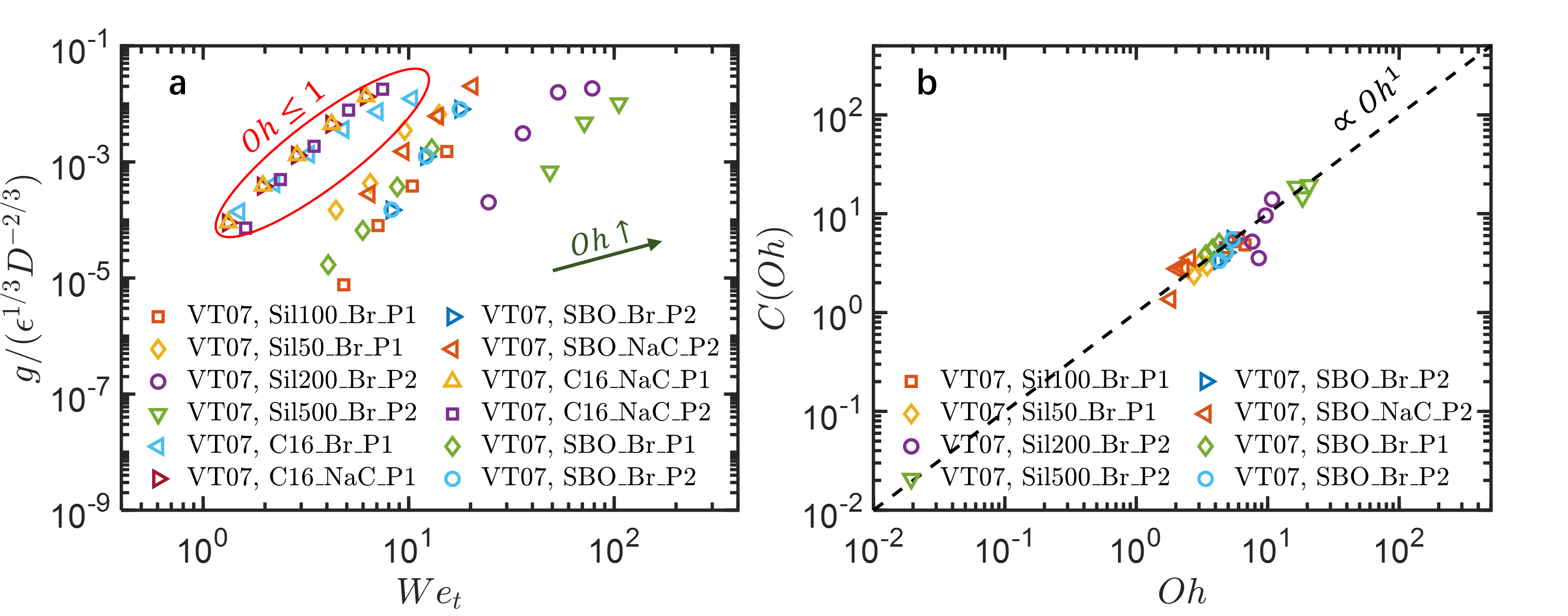}
    \caption{The effective Weber number coefficient, i.e. $C(Oh)$, versus $Oh$. (a) The breakup frequency normalized by the droplet-sized eddy turnover time versus Weber number. (b) $C(Oh)$ versus $Oh$ for the cases where $Oh>1$. The dashed line represents $C(Oh)=Oh$.}
    \label{fig:vankova_scale}
\end{figure*}

As for models based on the droplet-eddy collision frequency, by assuming the most efficient collision is by droplet-sized eddies, the breakup time should scale with the droplet sized eddy turnover time. According to the findings of \citet{vankova2007emulsification}, this model provided the best fit for their data. However, 
many other models have also been proposed \citep{tsouris1994breakage, wang2003novel, qi2020towards}. With appropriate fittings of their parameters, they may result in similar agreement with the experimental results so it is difficult to determine which model works better. 
In this study, instead of fitting many parameters for a given model, we decide to start from the data by \citet{vankova2007emulsification} and find out the simplest way to collapse all the results under one framework. 
To find out the possible functional form between $We_t$ and $Oh$, we first plot the normalized breakup frequency data versus the Weber number, as shown in panel a of Fig.\ref{fig:vankova_scale}. All the data with $Oh\leq 1$ falls onto the same master curve, since the viscosity effect is not dominant and Weber number is the control parameter. It is also evident that, for $Oh>1$, the data shifts to the right as $Oh$ increases. To make the data with $Oh>1$ falls onto the same curve, we assume the effective Weber number to be, $We_{t,\text{eff}}=We_{t}/C(Oh)$, where $C(Oh)$ is an unknown function of $Oh$. Then for each normalized breakup frequency lying outside the master curve, we calculate the corresponding $We_{t,\text{eff}}$ based on interpolation. Finally, we can calculate $C(Oh)$ based on the ratio between $We_{t,\text{eff}}$ and the its original $We_t$. 
%
% If $C(Oh)$ is a linear function of $Oh$, then all the data points will lie on a curve that is parallel to the dashed line in panel b of Fig.\ref{fig:vankova_scale}. Based on power law fitting, the scaling relation can be found to be $C(Oh)\sim Oh^{1.16}$, which is close to linear relation. 
Fig.\ref{fig:vankova_scale} shows the dependence of $C(Oh)$ versus $Oh$. The dashed line indicates a linear function, i.e. $C(Oh) \sim Oh$, and the least square fit of the data suggests $C(Oh)\sim Oh^{1.16}$. The nice collapse of the data against the dashed line suggests that a simple linear relationship is sufficient to describe $C(Oh)$.
In order to combine the data with $Oh\leq 1$ and that with $Oh>1$ together, we propose to use the simplest switching function, defined as $We_t/(1+Oh)$, which recovers the $We_t$ dependence as $Oh\rightarrow 0$, and $We_t/Oh$ as $Oh\rightarrow \infty$. 

Then we compare the data collapse in both the way suggested by \citet{vankova2007emulsification} against the one proposed above. 
In the upper panels (a and b) of Fig.\ref{fig:vankova_all}, the horizontal axis is defined as $We_t/(1+ c \cdot Oh^2 Re_d \sqrt{\rho_c/\rho_d} )$, which is based on Eq.\ref{eq:effectiveWe} with the fitting parameter, i.e. $c$, set as the same value shown in the paper. Although it indeed makes the data gathering closer, several data points still lie outside the curve. A further issue is that such a fitting parameter often varies within the literature, making it difficult to unify different datasets. 
As shown in the lower panels (c and d), using $We_t/(1+Oh)$, the data points collapses better than their corresponding upper panels (a and b). 
While the reason behind the improved data collapse using $We_t/(1+Oh)$ is not fully understood, it works the best at least for this dataset. Furthermore, it is important to point out that the exact functional form of $Oh$ is still an open question. \citet{pilch1987use} proposed an empirical power law of $Oh^{1.6}$. Further research is needed to explore and validate these assumptions and understand the underlying mechanisms.

Comparing panel c and d, it can be seen that normalizing $g$ with the eddy turnover frequency results in slightly better fitting compared to scaling it with the frequency modulated by the inner density or viscosity, $f(\mu_d)$ (Eq.\ref{eqn:fRe}). This observation suggests that models based on eddy collision frequency offer a better representation of the actual breakup process.
A recent study by \citet{qi2022fragmentation} provides experimental evidence indicating that the breakup is triggered by the collision of bubbles with intense eddy. A later work proposed that such intense collisions occur within a bubble/droplet sized eddy turnover time \citep{qi2023timescale}, lending further support to the scaling with eddy turnover frequency. These findings contribute to our understanding of the dominant mechanisms driving breakup and highlight the importance of considering the role of eddies in the process.

% \begin{figure}
%     \centering
%     \includegraphics[width=\linewidth]{Fig/vankova_fit_te.eps}
%     \caption{Normalized breakup frequency versus the effective Weber number. In this plot, the breakup frequency is normalized by the droplet size eddy turnover time; the effective Weber number is based on Eq.\ref{eq:effectiveWe}.}
%     \label{fig:vankova_fit_te}
% \end{figure}

% \begin{figure}
%     \centering
%     \includegraphics[width=\linewidth]{Fig/vankova_fit_tRe.eps}
%     \caption{Normalized breakup frequency versus the effective Weber number. In this plot, the breakup frequency is normalized by Eq.\ref{eqn:timeD} ($Re_d<1$) or droplet size eddy turnover time ($Re_d\geq 1$); the effective Weber number is based on Eq.\ref{eq:effectiveWe}.}
%     \label{fig:vankova_fit_tRe}
% \end{figure}

% \begin{figure}
%     \centering
%     \includegraphics[width=\linewidth]{Fig/vankova_our_te.eps}
%     \caption{Normalized breakup frequency versus the effective Weber number. In this plot, the breakup frequency is normalized by droplet size eddy turnover time; the new effective Weber number is defined as $We_t/(1+Oh)$.}
%     \label{fig:vankova_our_te}
% \end{figure}
% \begin{figure}
%     \centering
%     \includegraphics[width=\linewidth]{Fig/vankova_our_tRe.eps}
%     \caption{Normalized breakup frequency versus the effective Weber number. In this plot, the breakup frequency is normalized by Eq.\ref{eqn:timeD} ($Re_d<1$) or droplet size eddy turnover time ($Re_d\geq 1$); the new effective Weber number is defined as $We_t/(1+Oh)$.}
%     \label{fig:vankova_our_tRe}
% \end{figure}

\section{Data compilation}

Finally, we would like to compile the breakup frequency data from both bubble and droplet together to see if there is a unified framework that can describe the breakup dynamics of both.  
The datasets involve VZS18 \citep{vejrazka2018}, MML99 \citep{martinez1999breakup}, SJ15 \citep{solsvik2015single}, VT07\citep{vankova2007emulsification}, EAL04 \citep{eastwood2004breakup}, VMA22 \citep{vela2022memoryless}, HFSJ20 \citep{hero2020single}. Models include the ones by CT77 \citep{coulaloglou1977description}, QTN22 \citep{qi2022fragmentation}, QMN20 \citep{qi2020towards}, and WWJ03 \citep{wang2003novel}.
Note that the breakup frequency datasets from MML99, EAL04, VT07 and VMA22 are calculated based on fitting Eq.\ref{eq:gFit}; the breakup frequency data of SJ15 and HFSJ20 are re-calculated based on Eq.\ref{eq:gDef}. 

\begin{figure*}
    \centering
    \includegraphics[width=1.0\textwidth]{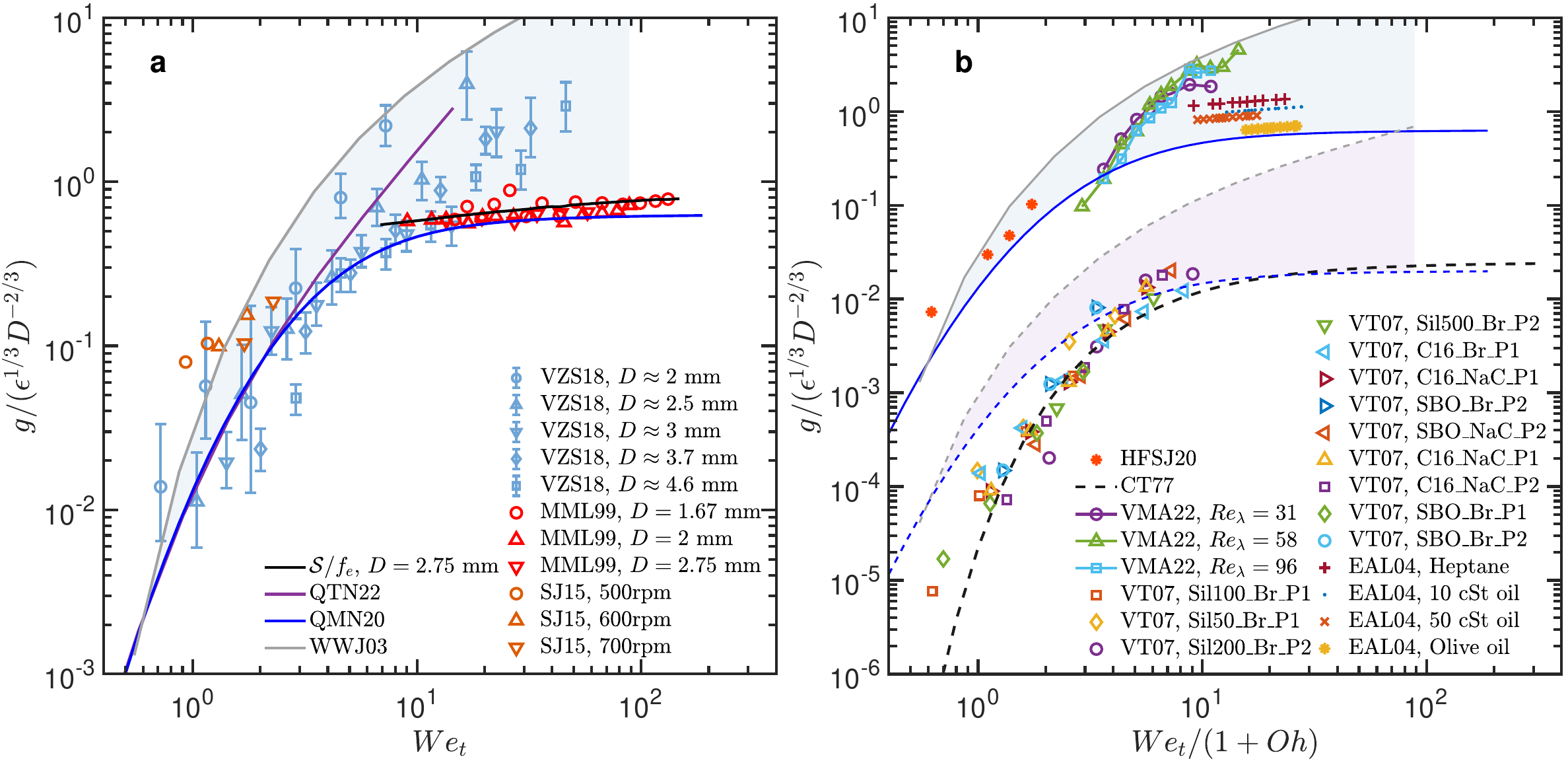}
    \caption{Breakup frequency of (a) bubbles and (b) droplets normalized by the eddy turnover frequency, $f_e=(\epsilon^{1/3}D^{-2/3})$, as a function of the key dimensionless number. The datasets that are compiled include VZS18 \citep{vejrazka2018}, MML99 \citep{martinez1999breakup}, SJ15 \citep{solsvik2015single}, VT07\citep{vankova2007emulsification}, EAL04 \citep{eastwood2004breakup}, VMA22 \citep{vela2022memoryless}, HFSJ20 \citep{hero2020single}. Models include the ones by CT77 \citep{coulaloglou1977description}, QTN22 \citep{qi2022fragmentation},QMN20 \citep{qi2020towards}, and WWJ03 \citep{wang2003novel}. 
    }
    \label{fig:breakfreq}
\end{figure*}

According to previous discussions, we choose to use bubble/droplet eddy turnover frequency to non-dimensionalize the breakup frequency and use $We_t$ as the horizontal axis for bubble's data and $We_t/(1+Oh)$ as the horizontal axis for droplet's data. 
As shown in Fig.\ref{fig:breakfreq}, all breakup frequency results for bubble are plotted in panel a, and the data for droplet is plotted in panel b. Both solid and dashed lines are based on breakup frequency models \citep{qi2020towards,qi2022fragmentation,wang2003novel,coulaloglou1977description} (dashed lines shown in panel b represent the breakup frequency times the density ratio, $\sqrt{\rho_d/\rho_c}$, based on Eq.\ref{equ:te_levich}); the experimental data is shown in symbols; the lines with symbols are from simulation results \citep{vela2022memoryless}.

The breakup frequency for bubbles can be effectively described by the QMN20 model (lower bound) proposed by \citep{wang2003novel} and the WWJ03 model (upper bound) proposed by \citep{qi2020towards}. When the Weber number ($We_t$) is less than 10, all the data exhibit a similar trend. However, when $We_t$ becomes much larger ($We_t \gg 10$), the asymptotic behavior of the breakup frequency remains uncertain. The data from Martinez-Bazan's experiments \citep{martinez1999breakup} suggests that the breakup frequency reaches a plateau, while the data from \citep{vejrazka2018} indicates that the breakup frequency continues to increase with increasing $We_t$, albeit with some scatter. 
In the study of breaking waves, a scaling law of the bubble size spectrum \citep{garrett2000connection,deane2002scale} was observed, 
\begin{equation}
    n(D) \sim D^{-10/3}, \quad \text{when $D_H \ll D \ll L$}
\end{equation}
where $D_H \sim \epsilon^{-2/5}(\sigma/(2\rho_c))^{3/5}$ is the Hinze scale \citep{deane2002scale}; $L$ is the integral length scale. It has been shown in many papers that such scaling could be related to the breakup frequency scaled as $g(D) \sim D^{-2/3}$, indicating that it might be more reasonable to observe an almost constant line when $We_t$ is large.
However, it is still an open question, and the actual asymptotic trend still needs further investigation.

Compared with bubble, the result of droplet could not be collapsed into one single line. For breakup frequency of large droplet ($\sim$1 mm) in experiment and in simulation \citep{hero2020single,vela2022memoryless,eastwood2004breakup}, they seem to lie in the same bounds as bubble; while the breakup frequency of small droplet ($\sim 50\mu$ m) is 2 order of magnitude smaller \citep{vankova2007emulsification}. This large difference might be due to the uncertainty of the residence time in homoginizer; it could also be possible that some unknown dependence on size that was not accounted for. As the size of droplet becomes small, the timescales for turbulence, capillary, and viscosity can be quite similar, which indicates that more than one effects might work together or against each other during the breakup process.

\section{Conclusion}

In this paper, we compiled and analyzed previous experimental datasets on fragmentation of bubbles and droplets by checking the experimental methods and processing the raw data, and evaluating different normalization options and characteristic scales. More importantly, it sheds light on a consistent framework for breakup frequency model of both bubble and droplet. The breakup timescale for HIT is well characterized by bubble/droplet-sized eddy turnover time; however, the relation to the shear timescale needs further investigations in anisotropic turbulence. It is found in this paper that by using the equivalent Weber number, i.e. $We_t/(1+Oh)$, both the bubble and droplet breakup frequency data can fall onto the same curve. However, it still remains unclear how to account for the inner density and viscosity for droplet breakup in turbulence. A further question is on the asymptotic behaviour for large inner viscosity: will the surface tension effect be completely gone or weakened gradually? 
The final question that we want to point out is the difference in decades between the droplet breakup data taken in the HIT setups (macroscale) and in the homoginizers (microscale), as shown in Fig.\ref{fig:breakfreq}. An exact explanation for that is lacking, but the possible reasons might be the uncertainty of residence time in homoginizer, the unknown dependence on the bubble/droplet size, or different breakup mechanisms \citep{masbernat2022prediction}.

In order to answer the above questions, more high-quality experimental data is needed, e.g. well resolved images from high speed camera, and facilities that can generate controllable large scale mean shear and turbulence. Large amount of data is required for convergence on the breakup frequency for different bubble/droplet sizes. 
Although lots of exciting findings have been published recently, many questions remain to be answered to provide a well constrained model for PBE and simulations.

\section*{Appendix}
\begin{figure*}
    \centering
    \includegraphics[width=\linewidth]{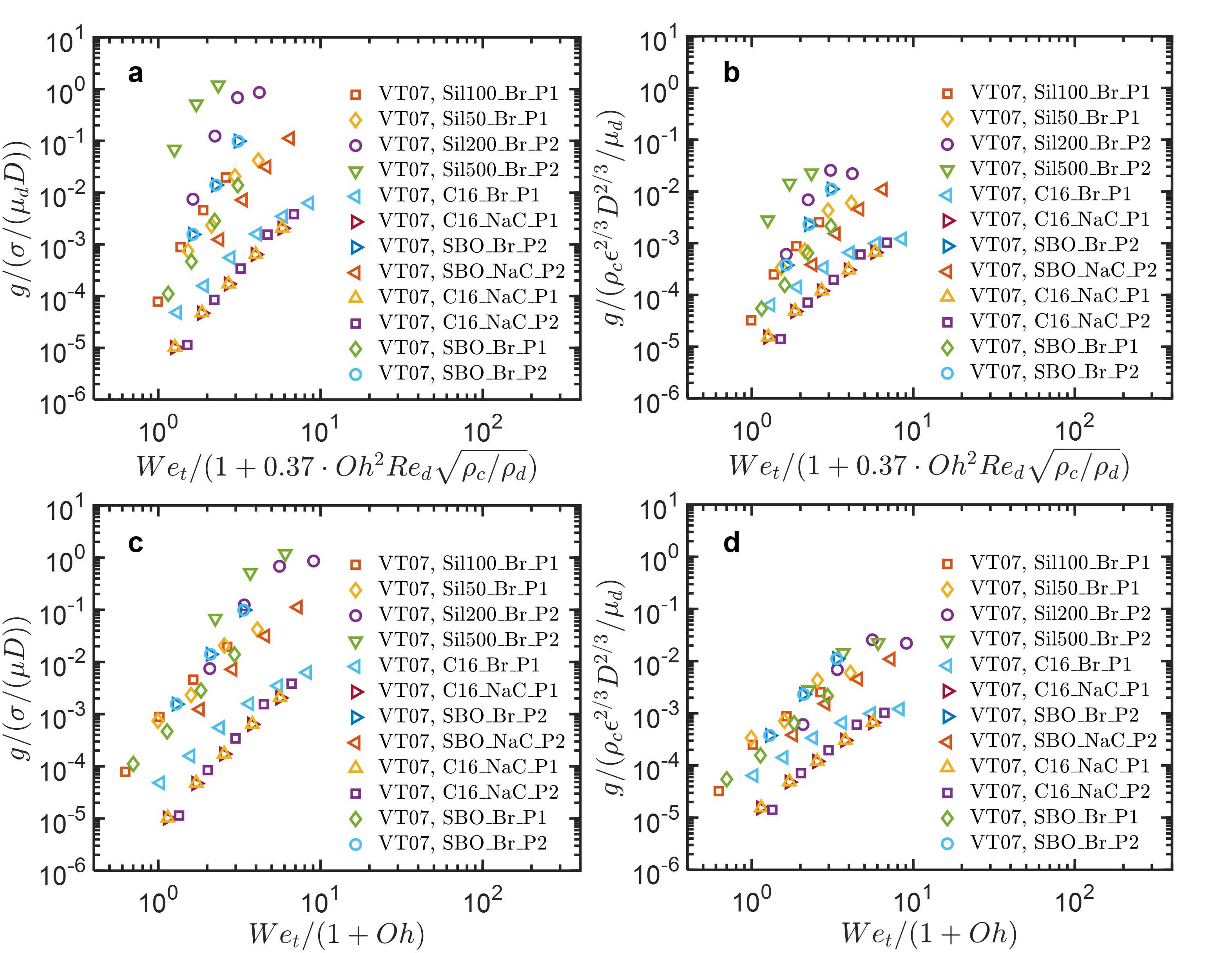}
    \caption{Normalized breakup frequency versus the effective Weber number. (a,c) the breakup frequency is normalized by the capillary time (Eq.\ref{equ:tc}). The horizontal axis in panel a is $We_t/(1+ c \cdot Oh^2 Re_d \sqrt{\rho_c/\rho_d} )$ with coefficient $c=0.37$ \citep{vankova2007emulsification}; while the horizontal axis in panel c is $We_t/(1+Oh)$. (b,d), the breakup frequency is normalized by Eq.\ref{equ:tv_levich} with $u_e\sim (\epsilon D)^{1/3}$.}
\end{figure*}

\printcredits

%% Loading bibliography style file
%\bibliographystyle{model1-num-names}
\bibliographystyle{cas-model2-names}
%\bibliographystyle{unsrt}

% Loading bibliography database
\bibliography{main_V2}

\begin{thebibliography}{49}
\expandafter\ifx\csname natexlab\endcsname\relax\def\natexlab#1{#1}\fi
\providecommand{\url}[1]{\texttt{#1}}
\providecommand{\href}[2]{#2}
\providecommand{\path}[1]{#1}
\providecommand{\DOIprefix}{doi:}
\providecommand{\ArXivprefix}{arXiv:}
\providecommand{\URLprefix}{URL: }
\providecommand{\Pubmedprefix}{pmid:}
\providecommand{\doi}[1]{\href{http://dx.doi.org/#1}{\path{#1}}}
\providecommand{\Pubmed}[1]{\href{pmid:#1}{\path{#1}}}
\providecommand{\bibinfo}[2]{#2}
\ifx\xfnm\relax \def\xfnm[#1]{\unskip,\space#1}\fi
%Type = Article
\bibitem[{Alopaeus et~al.(2002)Alopaeus, Koskinen, Keskinen and
  Majander}]{alopaeus2002simulation}
\bibinfo{author}{Alopaeus, V.}, \bibinfo{author}{Koskinen, J.},
  \bibinfo{author}{Keskinen, K.I.}, \bibinfo{author}{Majander, J.},
  \bibinfo{year}{2002}.
\newblock \bibinfo{title}{Simulation of the population balances for
  liquid--liquid systems in a nonideal stirred tank. part 2—parameter fitting
  and the use of the multiblock model for dense dispersions}.
\newblock \bibinfo{journal}{Chemical Engineering Science} \bibinfo{volume}{57},
  \bibinfo{pages}{1815--1825}.
%Type = Article
\bibitem[{Calabrese et~al.(1986)Calabrese, Chang and Dang}]{calabrese1986drop}
\bibinfo{author}{Calabrese, R.V.}, \bibinfo{author}{Chang, T.},
  \bibinfo{author}{Dang, P.}, \bibinfo{year}{1986}.
\newblock \bibinfo{title}{Drop breakup in turbulent stirred-tank contactors.
  {P}art {I}: Effect of dispersed-phase viscosity}.
\newblock \bibinfo{journal}{AIChE Journal} \bibinfo{volume}{32},
  \bibinfo{pages}{657--666}.
%Type = Article
\bibitem[{Chan et~al.(2021)Chan, Johnson, Moin and Urzay}]{chan2021turbulent}
\bibinfo{author}{Chan, W.H.R.}, \bibinfo{author}{Johnson, P.L.},
  \bibinfo{author}{Moin, P.}, \bibinfo{author}{Urzay, J.},
  \bibinfo{year}{2021}.
\newblock \bibinfo{title}{The turbulent bubble break-up cascade. part 2.
  numerical simulations of breaking waves}.
\newblock \bibinfo{journal}{Journal of Fluid Mechanics} \bibinfo{volume}{912}.
%Type = Article
\bibitem[{Coulaloglou and Tavlarides(1977)}]{coulaloglou1977description}
\bibinfo{author}{Coulaloglou, C.}, \bibinfo{author}{Tavlarides, L.L.},
  \bibinfo{year}{1977}.
\newblock \bibinfo{title}{Description of interaction processes in agitated
  liquid-liquid dispersions}.
\newblock \bibinfo{journal}{Chemical Engineering Science} \bibinfo{volume}{32},
  \bibinfo{pages}{1289--1297}.
%Type = Article
\bibitem[{Davies(1985)}]{davies1985drop}
\bibinfo{author}{Davies, J.}, \bibinfo{year}{1985}.
\newblock \bibinfo{title}{Drop sizes of emulsions related to turbulent energy
  dissipation rates}.
\newblock \bibinfo{journal}{Chemical Engineering Science} \bibinfo{volume}{40},
  \bibinfo{pages}{839--842}.
%Type = Article
\bibitem[{Deane and Stokes(2002)}]{deane2002scale}
\bibinfo{author}{Deane, G.}, \bibinfo{author}{Stokes, M.},
  \bibinfo{year}{2002}.
\newblock \bibinfo{title}{Scale dependence of bubble creation mechanisms in
  breaking waves}.
\newblock \bibinfo{journal}{Nature} \bibinfo{volume}{418},
  \bibinfo{pages}{839}.
%Type = Article
\bibitem[{Eastwood et~al.(2004)Eastwood, Armi and
  Lasheras}]{eastwood2004breakup}
\bibinfo{author}{Eastwood, C.D.}, \bibinfo{author}{Armi, L.},
  \bibinfo{author}{Lasheras, J.}, \bibinfo{year}{2004}.
\newblock \bibinfo{title}{The breakup of immiscible fluids in turbulent flows}.
\newblock \bibinfo{journal}{Journal of Fluid Mechanics} \bibinfo{volume}{502},
  \bibinfo{pages}{309--333}.
%Type = Article
\bibitem[{Garrett et~al.(2000)Garrett, Li and Farmer}]{garrett2000connection}
\bibinfo{author}{Garrett, C.}, \bibinfo{author}{Li, M.},
  \bibinfo{author}{Farmer, D.}, \bibinfo{year}{2000}.
\newblock \bibinfo{title}{The connection between bubble size spectra and energy
  dissipation rates in the upper ocean}.
\newblock \bibinfo{journal}{Journal of Physical Oceanography}
  \bibinfo{volume}{30}, \bibinfo{pages}{2163--2171}.
%Type = Article
\bibitem[{Gaylo et~al.(2023)Gaylo, Hendrickson and Yue}]{gaylo2023fundamental}
\bibinfo{author}{Gaylo, D.B.}, \bibinfo{author}{Hendrickson, K.},
  \bibinfo{author}{Yue, D.K.}, \bibinfo{year}{2023}.
\newblock \bibinfo{title}{Fundamental time scales of bubble fragmentation in
  homogeneous isotropic turbulence}.
\newblock \bibinfo{journal}{Journal of Fluid Mechanics} \bibinfo{volume}{962},
  \bibinfo{pages}{A25}.
%Type = Article
\bibitem[{Gourdon and Casamatta(1991)}]{gourdon1991influence}
\bibinfo{author}{Gourdon, C.}, \bibinfo{author}{Casamatta, G.},
  \bibinfo{year}{1991}.
\newblock \bibinfo{title}{Influence of mass transfer direction on the operation
  of a pulsed sieve-plate pilot column}.
\newblock \bibinfo{journal}{Chemical Engineering Science} \bibinfo{volume}{46},
  \bibinfo{pages}{2799--2808}.
%Type = Article
\bibitem[{Graham and Griffith(1973)}]{graham1973drop}
\bibinfo{author}{Graham, C.}, \bibinfo{author}{Griffith, P.},
  \bibinfo{year}{1973}.
\newblock \bibinfo{title}{Drop size distributions and heat transfer in dropwise
  condensation}.
\newblock \bibinfo{journal}{International Journal of Heat and Mass Transfer}
  \bibinfo{volume}{16}, \bibinfo{pages}{337--346}.
%Type = Article
\bibitem[{Gupta et~al.(2016)Gupta, Eral, Hatton and
  Doyle}]{gupta2016nanoemulsions}
\bibinfo{author}{Gupta, A.}, \bibinfo{author}{Eral, H.B.},
  \bibinfo{author}{Hatton, T.A.}, \bibinfo{author}{Doyle, P.S.},
  \bibinfo{year}{2016}.
\newblock \bibinfo{title}{Nanoemulsions: formation, properties and
  applications}.
\newblock \bibinfo{journal}{Soft Matter} \bibinfo{volume}{12},
  \bibinfo{pages}{2826--2841}.
%Type = Article
\bibitem[{H{\aa}kansson(2019)}]{haakansson2019emulsion}
\bibinfo{author}{H{\aa}kansson, A.}, \bibinfo{year}{2019}.
\newblock \bibinfo{title}{Emulsion formation by homogenization: Current
  understanding and future perspectives}.
\newblock \bibinfo{journal}{Annual Review of Food Science and Technology}
  \bibinfo{volume}{10}, \bibinfo{pages}{239--258}.
%Type = Article
\bibitem[{H{\aa}kansson(2020a)}]{haakansson2020experimental}
\bibinfo{author}{H{\aa}kansson, A.}, \bibinfo{year}{2020}a.
\newblock \bibinfo{title}{Experimental methods for measuring the breakup
  frequency in turbulent emulsification: A critical review}.
\newblock \bibinfo{journal}{ChemEngineering} \bibinfo{volume}{4},
  \bibinfo{pages}{52}.
%Type = Article
\bibitem[{H{\aa}kansson(2020b)}]{haakansson2020validity}
\bibinfo{author}{H{\aa}kansson, A.}, \bibinfo{year}{2020}b.
\newblock \bibinfo{title}{On the validity of different methods to estimate
  breakup frequency from single drop experiments}.
\newblock \bibinfo{journal}{Chemical Engineering Science}
  \bibinfo{volume}{227}, \bibinfo{pages}{115908}.
%Type = Article
\bibitem[{Han{\v{c}}il and Rod(1988)}]{hanvcil1988break}
\bibinfo{author}{Han{\v{c}}il, V.}, \bibinfo{author}{Rod, V.},
  \bibinfo{year}{1988}.
\newblock \bibinfo{title}{Break-up of a drop in a stirred tank}.
\newblock \bibinfo{journal}{Chemical Engineering and Processing: Process
  Intensification} \bibinfo{volume}{23}, \bibinfo{pages}{189--193}.
%Type = Article
\bibitem[{Her{\o} et~al.(2020)Her{\o}, La~Forgia, Solsvik and
  Jakobsen}]{hero2020single}
\bibinfo{author}{Her{\o}, E.H.}, \bibinfo{author}{La~Forgia, N.},
  \bibinfo{author}{Solsvik, J.}, \bibinfo{author}{Jakobsen, H.A.},
  \bibinfo{year}{2020}.
\newblock \bibinfo{title}{Single drop breakage in turbulent flow: Statistical
  data analysis}.
\newblock \bibinfo{journal}{Chemical Engineering Science: X}
  \bibinfo{volume}{8}, \bibinfo{pages}{100082}.
%Type = Article
\bibitem[{Hinze(1955)}]{hinze1955fundamentals}
\bibinfo{author}{Hinze, J.}, \bibinfo{year}{1955}.
\newblock \bibinfo{title}{Fundamentals of the hydrodynamic mechanism of
  splitting in dispersion processes}.
\newblock \bibinfo{journal}{AIChE Journal} \bibinfo{volume}{1},
  \bibinfo{pages}{289--295}.
%Type = Article
\bibitem[{Hounslow and Ni(2004)}]{hounslow2004population}
\bibinfo{author}{Hounslow, M.}, \bibinfo{author}{Ni, X.}, \bibinfo{year}{2004}.
\newblock \bibinfo{title}{Population balance modelling of droplet coalescence
  and break-up in an oscillatory baffled reactor}.
\newblock \bibinfo{journal}{Chemical Engineering Science} \bibinfo{volume}{59},
  \bibinfo{pages}{819--828}.
%Type = Inproceedings
\bibitem[{Kolmogorov(1949)}]{kolmogorov1949breakage}
\bibinfo{author}{Kolmogorov, A.}, \bibinfo{year}{1949}.
\newblock \bibinfo{title}{On the breakage of drops in a turbulent flow}, in:
  \bibinfo{booktitle}{Dokl. Akad. Navk. SSSR}, pp. \bibinfo{pages}{825--828}.
%Type = Article
\bibitem[{Lalanne et~al.(2019)Lalanne, Masbernat and Risso}]{lalanne2019model}
\bibinfo{author}{Lalanne, B.}, \bibinfo{author}{Masbernat, O.},
  \bibinfo{author}{Risso, F.}, \bibinfo{year}{2019}.
\newblock \bibinfo{title}{A model for drop and bubble breakup frequency based
  on turbulence spectra}.
\newblock \bibinfo{journal}{AIChE Journal} \bibinfo{volume}{65},
  \bibinfo{pages}{347--359}.
%Type = Article
\bibitem[{Lasheras et~al.(2002)Lasheras, Eastwood, Mart{\i}nez-Baz{\'a}n and
  Montanes}]{lasheras2002review}
\bibinfo{author}{Lasheras, J.C.}, \bibinfo{author}{Eastwood, C.},
  \bibinfo{author}{Mart{\i}nez-Baz{\'a}n, C.}, \bibinfo{author}{Montanes, J.},
  \bibinfo{year}{2002}.
\newblock \bibinfo{title}{A review of statistical models for the break-up of an
  immiscible fluid immersed into a fully developed turbulent flow}.
\newblock \bibinfo{journal}{International Journal of Multiphase Flow}
  \bibinfo{volume}{28}, \bibinfo{pages}{247--278}.
%Type = Article
\bibitem[{Lehr et~al.(2002)Lehr, Millies and Mewes}]{lehr2002}
\bibinfo{author}{Lehr, F.}, \bibinfo{author}{Millies, M.},
  \bibinfo{author}{Mewes, D.}, \bibinfo{year}{2002}.
\newblock \bibinfo{title}{Bubble-size distributions and flow fields in bubble
  columns}.
\newblock \bibinfo{journal}{AIChE Journal} \bibinfo{volume}{48},
  \bibinfo{pages}{2426--2443}.
%Type = Book
\bibitem[{Levich(1962)}]{levich1962physicochemical}
\bibinfo{author}{Levich, V.G.}, \bibinfo{year}{1962}.
\newblock \bibinfo{title}{Physicochemical hydrodynamics}.
\newblock \bibinfo{publisher}{Prentice-Hall Inc.}
%Type = Article
\bibitem[{Liao and Lucas(2009)}]{liao2009literature}
\bibinfo{author}{Liao, Y.}, \bibinfo{author}{Lucas, D.}, \bibinfo{year}{2009}.
\newblock \bibinfo{title}{A literature review of theoretical models for drop
  and bubble breakup in turbulent dispersions}.
\newblock \bibinfo{journal}{Chemical Engineering Science} \bibinfo{volume}{64},
  \bibinfo{pages}{3389--3406}.
%Type = Book
\bibitem[{Marchisio and Fox(2013)}]{marchisio2013computational}
\bibinfo{author}{Marchisio, D.L.}, \bibinfo{author}{Fox, R.O.},
  \bibinfo{year}{2013}.
\newblock \bibinfo{title}{Computational models for polydisperse particulate and
  multiphase systems}.
\newblock \bibinfo{publisher}{Cambridge University Press}.
%Type = Article
\bibitem[{Marchisio et~al.(2003)Marchisio, Vigil and
  Fox}]{marchisio2003quadrature}
\bibinfo{author}{Marchisio, D.L.}, \bibinfo{author}{Vigil, R.D.},
  \bibinfo{author}{Fox, R.O.}, \bibinfo{year}{2003}.
\newblock \bibinfo{title}{Quadrature method of moments for
  aggregation--breakage processes}.
\newblock \bibinfo{journal}{Journal of Colloid and Interface Science}
  \bibinfo{volume}{258}, \bibinfo{pages}{322--334}.
%Type = Article
\bibitem[{Mart{\'i}nez-Baz{\'a}n et~al.(1999)Mart{\'i}nez-Baz{\'a}n, Montanes
  and Lasheras}]{martinez1999breakup}
\bibinfo{author}{Mart{\'i}nez-Baz{\'a}n, C.}, \bibinfo{author}{Montanes, J.},
  \bibinfo{author}{Lasheras, J.C.}, \bibinfo{year}{1999}.
\newblock \bibinfo{title}{On the breakup of an air bubble injected into a fully
  developed turbulent flow. {P}art 1. breakup frequency}.
\newblock \bibinfo{journal}{Journal of Fluid Mechanics} \bibinfo{volume}{401},
  \bibinfo{pages}{157--182}.
%Type = Article
\bibitem[{Mart{\'i}nez-Baz{\'a}n et~al.(2010)Mart{\'i}nez-Baz{\'a}n,
  RODR{\'I}GUEZ-RODR{\'I}GUEZ, Deane, Montanes and
  Lasheras}]{martinez2010considerations}
\bibinfo{author}{Mart{\'i}nez-Baz{\'a}n, C.},
  \bibinfo{author}{RODR{\'I}GUEZ-RODR{\'I}GUEZ, J.}, \bibinfo{author}{Deane,
  G.}, \bibinfo{author}{Montanes, J.}, \bibinfo{author}{Lasheras, J.},
  \bibinfo{year}{2010}.
\newblock \bibinfo{title}{Considerations on bubble fragmentation models}.
\newblock \bibinfo{journal}{Journal of Fluid Mechanics} \bibinfo{volume}{661},
  \bibinfo{pages}{159--177}.
%Type = Article
\bibitem[{Masbernat et~al.(2022)Masbernat, Risso, Lalanne, Bugeat and
  Berton}]{masbernat2022prediction}
\bibinfo{author}{Masbernat, O.}, \bibinfo{author}{Risso, F.},
  \bibinfo{author}{Lalanne, B.}, \bibinfo{author}{Bugeat, S.},
  \bibinfo{author}{Berton, M.}, \bibinfo{year}{2022}.
\newblock \bibinfo{title}{Prediction of size distribution in dairy cream
  homogenization}.
\newblock \bibinfo{journal}{Journal of Food Engineering} \bibinfo{volume}{324},
  \bibinfo{pages}{110973}.
%Type = Article
\bibitem[{Mcclements(2007)}]{mcclements2007critical}
\bibinfo{author}{Mcclements, D.J.}, \bibinfo{year}{2007}.
\newblock \bibinfo{title}{Critical review of techniques and methodologies for
  characterization of emulsion stability}.
\newblock \bibinfo{journal}{Critical Reviews in Food Science and Nutrition}
  \bibinfo{volume}{47}, \bibinfo{pages}{611--649}.
%Type = Article
\bibitem[{O’Rourke and MacLoughlin(2010)}]{o2010study}
\bibinfo{author}{O’Rourke, A.M.}, \bibinfo{author}{MacLoughlin, P.},
  \bibinfo{year}{2010}.
\newblock \bibinfo{title}{A study of drop breakage in lean dispersions using
  the inverse-problem method}.
\newblock \bibinfo{journal}{Chemical Engineering Science} \bibinfo{volume}{65},
  \bibinfo{pages}{3681--3694}.
%Type = Article
\bibitem[{Pilch and Erdman(1987)}]{pilch1987use}
\bibinfo{author}{Pilch, M.}, \bibinfo{author}{Erdman, C.},
  \bibinfo{year}{1987}.
\newblock \bibinfo{title}{Use of breakup time data and velocity history data to
  predict the maximum size of stable fragments for acceleration-induced breakup
  of a liquid drop}.
\newblock \bibinfo{journal}{International Journal of Multiphase Flow}
  \bibinfo{volume}{13}, \bibinfo{pages}{741--757}.
%Type = Article
\bibitem[{Prince and Blanch(1990)}]{prince1990bubble}
\bibinfo{author}{Prince, M.J.}, \bibinfo{author}{Blanch, H.W.},
  \bibinfo{year}{1990}.
\newblock \bibinfo{title}{Bubble coalescence and break-up in air-sparged bubble
  columns}.
\newblock \bibinfo{journal}{AIChE Journal} \bibinfo{volume}{36},
  \bibinfo{pages}{1485--1499}.
%Type = Article
\bibitem[{Qi et~al.(2020)Qi, Masuk and Ni}]{qi2020towards}
\bibinfo{author}{Qi, Y.}, \bibinfo{author}{Masuk, A.U.M.}, \bibinfo{author}{Ni,
  R.}, \bibinfo{year}{2020}.
\newblock \bibinfo{title}{Towards a model of bubble breakup in turbulence
  through experimental constraints}.
\newblock \bibinfo{journal}{International Journal of Multiphase Flow}
  \bibinfo{volume}{132}, \bibinfo{pages}{103397}.
%Type = Article
\bibitem[{Qi et~al.(2022)Qi, Tan, Corbitt, Urbanik, Salibindla and
  Ni}]{qi2022fragmentation}
\bibinfo{author}{Qi, Y.}, \bibinfo{author}{Tan, S.}, \bibinfo{author}{Corbitt,
  N.}, \bibinfo{author}{Urbanik, C.}, \bibinfo{author}{Salibindla, A.K.},
  \bibinfo{author}{Ni, R.}, \bibinfo{year}{2022}.
\newblock \bibinfo{title}{Fragmentation in turbulence by small eddies}.
\newblock \bibinfo{journal}{Nature Communications} \bibinfo{volume}{13},
  \bibinfo{pages}{1--8}.
%Type = Article
\bibitem[{Ravichandar et~al.(2023)Ravichandar, Olsen and
  Vigil}]{ravichandar2023turbulent}
\bibinfo{author}{Ravichandar, K.}, \bibinfo{author}{Olsen, M.G.},
  \bibinfo{author}{Vigil, R.D.}, \bibinfo{year}{2023}.
\newblock \bibinfo{title}{Turbulent droplet breakage probability: Analysis of
  fitting parameters for two commonly used models}.
\newblock \bibinfo{journal}{Chemical Engineering Science}
  \bibinfo{volume}{266}, \bibinfo{pages}{118311}.
%Type = Article
\bibitem[{Shiea et~al.(2020)Shiea, Buffo, Vanni and
  Marchisio}]{shiea2020numerical}
\bibinfo{author}{Shiea, M.}, \bibinfo{author}{Buffo, A.},
  \bibinfo{author}{Vanni, M.}, \bibinfo{author}{Marchisio, D.},
  \bibinfo{year}{2020}.
\newblock \bibinfo{title}{Numerical methods for the solution of population
  balance equations coupled with computational fluid dynamics}.
\newblock \bibinfo{journal}{Annual Review of Chemical and Biomolecular
  Engineering} \bibinfo{volume}{11}, \bibinfo{pages}{339--366}.
%Type = Article
\bibitem[{Solsvik and Jakobsen(2015)}]{solsvik2015single}
\bibinfo{author}{Solsvik, J.}, \bibinfo{author}{Jakobsen, H.A.},
  \bibinfo{year}{2015}.
\newblock \bibinfo{title}{Single air bubble breakup experiments in stirred
  water tank}.
\newblock \bibinfo{journal}{International Journal of Chemical Reactor
  Engineering} \bibinfo{volume}{13}, \bibinfo{pages}{477--491}.
%Type = Article
\bibitem[{Tcholakova et~al.(2007)Tcholakova, Vankova, Denkov and
  Danner}]{tcholakova2007emulsificationDaughter}
\bibinfo{author}{Tcholakova, S.}, \bibinfo{author}{Vankova, N.},
  \bibinfo{author}{Denkov, N.D.}, \bibinfo{author}{Danner, T.},
  \bibinfo{year}{2007}.
\newblock \bibinfo{title}{Emulsification in turbulent flow:: 3. daughter
  drop-size distribution}.
\newblock \bibinfo{journal}{Journal of Colloid and Interface Science}
  \bibinfo{volume}{310}, \bibinfo{pages}{570--589}.
%Type = Article
\bibitem[{Tsouris and Tavlarides(1994)}]{tsouris1994breakage}
\bibinfo{author}{Tsouris, C.}, \bibinfo{author}{Tavlarides, L.L.},
  \bibinfo{year}{1994}.
\newblock \bibinfo{title}{Breakage and coalescence models for drops in
  turbulent dispersions}.
\newblock \bibinfo{journal}{AIChE Journal} \bibinfo{volume}{40},
  \bibinfo{pages}{395--406}.
%Type = Article
\bibitem[{Vankova et~al.(2007a)Vankova, Tcholakova, Denkov, Ivanov, Vulchev and
  Danner}]{vankova2007emulsificationSize}
\bibinfo{author}{Vankova, N.}, \bibinfo{author}{Tcholakova, S.},
  \bibinfo{author}{Denkov, N.D.}, \bibinfo{author}{Ivanov, I.B.},
  \bibinfo{author}{Vulchev, V.D.}, \bibinfo{author}{Danner, T.},
  \bibinfo{year}{2007}a.
\newblock \bibinfo{title}{Emulsification in turbulent flow: 1. mean and maximum
  drop diameters in inertial and viscous regimes}.
\newblock \bibinfo{journal}{Journal of Colloid and Interface Science}
  \bibinfo{volume}{312}, \bibinfo{pages}{363--380}.
%Type = Article
\bibitem[{Vankova et~al.(2007b)Vankova, Tcholakova, Denkov, Vulchev and
  Danner}]{vankova2007emulsification}
\bibinfo{author}{Vankova, N.}, \bibinfo{author}{Tcholakova, S.},
  \bibinfo{author}{Denkov, N.D.}, \bibinfo{author}{Vulchev, V.D.},
  \bibinfo{author}{Danner, T.}, \bibinfo{year}{2007}b.
\newblock \bibinfo{title}{Emulsification in turbulent flow: 2. breakage rate
  constants}.
\newblock \bibinfo{journal}{Journal of Colloid and Interface Science}
  \bibinfo{volume}{313}, \bibinfo{pages}{612--629}.
%Type = Article
\bibitem[{Vejra{\v{z}}ka et~al.(2018)Vejra{\v{z}}ka, Zedn{\'\i}kov{\'a} and
  Stanovsk{\'y}}]{vejrazka2018}
\bibinfo{author}{Vejra{\v{z}}ka, J.}, \bibinfo{author}{Zedn{\'\i}kov{\'a}, M.},
  \bibinfo{author}{Stanovsk{\'y}, P.}, \bibinfo{year}{2018}.
\newblock \bibinfo{title}{Experiments on breakup of bubbles in a turbulent
  flow}.
\newblock \bibinfo{journal}{AIChE Journal} \bibinfo{volume}{64},
  \bibinfo{pages}{740--757}.
%Type = Article
\bibitem[{Vela-Mart{\'\i}n and Avila(2022)}]{vela2022memoryless}
\bibinfo{author}{Vela-Mart{\'\i}n, A.}, \bibinfo{author}{Avila, M.},
  \bibinfo{year}{2022}.
\newblock \bibinfo{title}{Memoryless drop breakup in turbulence}.
\newblock \bibinfo{journal}{Science Advances} \bibinfo{volume}{8},
  \bibinfo{pages}{eabp9561}.
%Type = Article
\bibitem[{Verschoof et~al.(2016)Verschoof, Van Der~Veen, Sun and
  Lohse}]{verschoof2016bubble}
\bibinfo{author}{Verschoof, R.A.}, \bibinfo{author}{Van Der~Veen, R.C.},
  \bibinfo{author}{Sun, C.}, \bibinfo{author}{Lohse, D.}, \bibinfo{year}{2016}.
\newblock \bibinfo{title}{Bubble drag reduction requires large bubbles}.
\newblock \bibinfo{journal}{Physical Review Letters} \bibinfo{volume}{117},
  \bibinfo{pages}{104502}.
%Type = Article
\bibitem[{Wang et~al.(2003)Wang, Wang and Jin}]{wang2003novel}
\bibinfo{author}{Wang, T.}, \bibinfo{author}{Wang, J.}, \bibinfo{author}{Jin,
  Y.}, \bibinfo{year}{2003}.
\newblock \bibinfo{title}{A novel theoretical breakup kernel function for
  bubbles/droplets in a turbulent flow}.
\newblock \bibinfo{journal}{Chemical Engineering Science} \bibinfo{volume}{58},
  \bibinfo{pages}{4629--4637}.
%Type = Book
\bibitem[{Williams(1985)}]{williams1985combustion}
\bibinfo{author}{Williams, F.A.}, \bibinfo{year}{1985}.
\newblock \bibinfo{title}{Combustion theory}.
\newblock \bibinfo{edition}{2nd} ed., \bibinfo{publisher}{CRC Press}.
%Type = Article
\bibitem[{Yinghe~Qi(2023)}]{qi2023timescale}
\bibinfo{author}{Yinghe~Qi, ..., R.N.}, \bibinfo{year}{2023}.
\newblock \bibinfo{title}{Multiple time scales in bubble breakup in
  turbulence}.
\newblock \bibinfo{journal}{Submitted} .

\end{thebibliography}

\end{document}